# Reversible switching of room temperature ferromagnetism in the $CeO_2$-Co nanoparticles

J. Sacanell[1,a,b], M. A. Paulin[1,2,b], V. Ferrari[1], G. Garbarino[3], A. G. Leyva[1,4]

[1] *Departamento de Física de la Materia Condensada, Centro Atómico Constituyentes, CNEA, Av. Gral. Paz 1499, Buenos Aires, Argentina.*

[2] *Departamento de Física, FCEN, Universidad de Buenos Aires, Buenos Aires, Argentina.*

[3] *European Synchrotron Radiation Facility, BP 220, F-38043, Grenoble Cedex, France*

[4] *Escuela de Ciencia y Tecnología, UNSAM, San Martín, Buenos Aires, Argentina.*

We investigated the reversible ferromagnetic (FM) behavior of pure and Co doped $CeO_2$ nanopowders. The as-sintered samples displayed an increasing paramagnetic contribution upon Co doping. Room temperature FM is obtained simply by performing thermal treatments in vacuum at temperatures as low as 500ºC and it can be switched off by performing thermal treatments in oxidizing conditions. The FM contribution is enhanced as we increase the time of the thermal treatment in vacuum. Those systematic experiments establish a direct relation between ferromagnetism and oxygen vacancies and open a path for developing materials with tailored properties.

---

[a] Author to whom correspondence should be addressed. Electronic mail: sacanell@cnea.gov.ar

[b] J. Sacanell and M. A. Paulin contributed equally to this work.

Diluted Magnetic Oxides[1] (DMOs) have been the focus of extensive research in the last years, due to the growing interest in finding materials for the emerging technology of spintronics[2]. The intriguing occurrence of room temperature ferromagnetism (RT-FM) in typically non-magnetic oxides under certain conditions, have also attracted much attention in the field of condensed matter physics and there is major consensus that point defects like oxygen vacancies play a fundamental role in the emergence of this phenomenon[1,3,4].

Among the various oxides displaying this particular behavior is $CeO_2$, which in bulk is a diamagnetic insulator with proven capability as an oxygen ion conductor when doped with trivalent Sm or Gd ions[5]. $CeO_2$-based systems have been thoroughly studied after the discovery of RT-FM on both pure[6,7,8,9,10] and slightly doped[11,12,13,14] samples. A large amount of effort has been devoted to the study of thin films[3,12,13,15,16,17,18,19] (see also references in [1]) while few works have focused on powdered samples[8,9,10,11,14]. The use of powders is advantageous because the typically low magnetic signals of these systems can easily be enhanced just by increasing the quantity of the sample. In addition, the surface to volume ratio of the sample can be adjusted simply by performing adequate thermal treatments and this is particularly convenient as the surface is usually the preferred location of point defects.

In several works devoted to RT-FM in $CeO_2$, the high mobility of oxygen, and the consequent facility to create oxygen vacancies, is suggested as the cause for this peculiar behavior as this process would promote the presence of magnetic $Ce^{3+}$ ions[3,15]. However, it is still not clear whether a small amount of magnetic ions can give rise to long range interactions even in slightly doped samples. In any case, this point has not yet been established beyond doubt[8] and even the presence of impurities[20] or clustering of the dopant[21] has not been completely discarded as one possible answer for the utterly unexpected magnetic behavior reported in the literature.

As magnetic signals in DMOs are typically small, extreme care must be taken during the synthesis procedure and handling of the samples to avoid contamination, and it is also important to carry out methodological studies to discard the possible influence of spurious magnetic impurities. In this work we performed a systematic study of the magnetic properties of $CeO_2$ nanopowders with particular focus on the influence of oxygen vacancies in both pure and doped samples. We show that an annealing process in vacuum can induce RT-FM in doped samples and paramagnetic (PM) behavior in pure samples. In all cases, this magnetic behavior can be erased by re-oxidation, ruling out segregation as responsible for the magnetic signal. The use of nanostructures allows us to enhance the surface to volume ratio, thus increasing the relative concentration of defects compared to bulk samples.

Pure and Co-doped $CeO_2$ powders were synthesized by the Liquid-Mix method using 99.99% $Ce(NO_3):6H_2O$ and Cobalt(II) nitrate hexahydrate as reagents. All samples were calcined in air at 300ºC. Additional thermal treatments were performed at 500ºC, both in air at atmospheric pressure and in vacuum, using a mechanical pump to reach a pressure of around $4.10^{-2}$ mbar, with dwell times between 4 and 30 hs. Samples were labeled with a two-number code denoting the atomic percentage of Co ($x$), and the dwell time of the vacuum treatment (*dwt*) in hours, namely *x-dwt*. Magnetic measurements were performed in a Versalab<sup>TM</sup> VSM from Quantum Design. X-ray diffraction data was obtained at the ID27 beamline of the European Synchrotron Radiation Facility.

The X-ray diffraction analysis of our samples, indicate a fluorite structure with no secondary phases[22]. The crystallite sizes of our samples are around 7 nm, obtained by the Scherrer equation. In the samples doped with more than 15% of Co, we observed some impurity peaks which can be indexed with cubic $Co_3O_4$ but they never overpass 1% in weight fraction.

In figure 1(a) we show the magnetic measurements at 400K for the samples synthesized in air for different doping concentrations. The behavior of the samples ranges from diamagnetic for the pure $CeO_2$ to an increasing PM signal upon doping. The magnetic component corresponding to the sample holder was previously measured and subtracted from all the presented data. Figure 1(b) shows the magnetic susceptibility of the as-sintered samples as a function of temperature. We see that the magnetic susceptibility of the pure sample is negative and almost temperature independent while the one corresponding to the doped samples displays the increasing PM tendency upon doping.

As we are interested in the role played by oxygen vacancies, we performed a series of thermal treatments in vacuum in order to de-oxygenate the samples. The color of the as-sintered samples change from yellow in the un-doped samples, to green in the doped ones. A first look at the samples treated in vacuum evidenced a darkening that strengths when increasing the treatment time, possibly indicating changes in the optical band-gap and in turn, the electronic structure of the system.

Figure 2 (a) and (b) show the M vs H dependence at 400K of samples with *dwt*=4 hs. The system displays a significant change which evidences the crucial role played by oxygen vacancies. Magnetization measurements of de-oxygenated samples, show a progressive evolution from a PM behavior for the low doped samples, to a FM response for samples with more than 9% of Co content. In fact, the magnetization of the doped samples is the result of the superposition of two components: a PM one and a FM one.

In figure 2 (c) we show measurements for samples with 9% of Co, with incremental *dwt*. In the samples treated in vacuum, the FM contribution is enhanced when *dwt* is increased. This result confirms the direct relation between the appearance of oxygen vacancies and the emergence of ferromagnetism. In the following, we present a separate analysis of the PM and the FM contributions to the magnetization.

The PM contribution to the magnetic susceptibility follows a Curie-Weiss law according to[23] $\chi = Np/[3k_B(T-\theta)]$, with $N$ the number density, $p$ the effective number of Bohr magnetons and $\theta$ the Curie-Weiss constant. In all cases, a negative $\theta$ was obtained of around -20 K. Non-reduced samples present $p \sim 3$, which suggests a magnetic contribution originated mainly by $Co^{2+}$ ions. In fact, the cobalt reagent and the low temperatures used during the synthesis procedure, is consistent with this picture. In the samples treated in vacuum for 4 hs, $p$ increases to about 4, indicating either a $Co^{3+}$ prominence or the presence of $Co^{2+}$ with a larger quantity (compared with non reduced samples) of $Ce^{3+}$ ions. Although we cannot rule out any of them, a strong suggestion in favor of the latter is that $CeO_2$ systems are very good oxygen ion conductors in presence of oxygen vacancies, a fact related to the facility of Ce ions to change from of $Ce^{4+}$ to $Ce^{3+}$. In addition, recent calculations[24] showed that oxygen vacancies tend to group near cobalt ions favoring the presence of $Co^{2+}$ and $Ce^{3+}$ ions. No significant changes are observed for longer vacuum treatments[25].

In the samples with $x$=15%, we observed the presence of $Co_3O_4$ impurities. $Co_3O_4$ is PM as measured in our synthesized bulk samples and it is no possible to magnetically individualize this contribution and separate it from that of Co as dopant. However, as we observed only 1% of this impurity, the correction to the overall susceptibility should be minor. For smaller Co concentrations, the correction becomes insignificant[26]. In addition, it is important to notice that the magnetic susceptibility of the pure sample with *dwt* = 4hs, is one order of magnitude below the one corresponding to doped samples. Thus the change from *p*~3 to *p*~4 is consistent with the appearance of some $Ce^{3+}$ ions during the vacuum treatment.

In figure 3 we present $M_{sat}$ (namely, the saturation of the FM component of the magnetization) as a function of doping for the samples treated in vacuum for 4hs and as a function of *dwt* for samples with 9% of Co (figure 3 (a) and (b), respectively). The nonlinear growth of $M_{sat}$ versus doping is consistent with an enhancement of the FM interaction, possibly related to the reduction of the distance between Co ions. In addition, the observed increment of $M_{sat}$ versus *dwt* indicates that an increase in the concentration of oxygen vacancies is essential to explain the appearance of ferromagnetism. This result is in agreement with previous theoretical and experimental works[11,14,17,24] suggesting that oxygen vacancies enhance FM behavior in the $CeO_2$-Co system.

Taking into account that the change in the magnetic properties only appears after thermal treatments, a doubt could be raised on the accidental incorporation of spurious magnetic impurities. In order to rule out this possibility we have oxidated sample *9-4* by performing an additional thermal treatment at 500ºC in air for 4 hs. As shown in figure 2 (d), the induced FM behavior is reversible, which unambiguously establishes the relation of ferromagnetism with the presence of oxygen vacancies. In addition, the colors of the samples return to the original yellow (or green) for the pure (or doped) samples. No evidence was observed on our X-ray data regarding the presence of metallic cobalt; however, as a very small amount of it could give rise to a significant signal this is a fact that deserves further study.

We have also tried to induce ferromagnetism on the pure sample by performing thermal treatments in vacuum. We found that an extensive thermal treatment of *dwt*=30 hs can only induce a PM dependence. We observed only a small relative change of the magnetic signal when increasing the dwell time of the treatment from 10 to 30 hs which evidences the futility of further increasing *dwt* and strongly suggests that we have reached an equilibrium state beyond which no changes are observed[27]. Detailed investigation on the effect of pressure of the vacuum treatment and alternative de-oxygenation procedures are currently in progress.

In summary, we performed a systematic investigation of the magnetic properties of Co-doped $CeO_2$ nanopowders. We observed that the as-sintered samples display a PM behavior (diamagnetic for non doped samples) and RT-FM was obtained by performing thermal treatments in vacuum. RT-FM is enhanced as doping is increased and it can be erased by performing thermal treatments in air showing a reversible behavior. These facts strongly suggest that oxygen vacancies are one of the main reasons for the appearance of RT-FM in the $CeO_2$-Co system. In all cases, an intrinsic PM component was observed superimposed to the FM one, indicating the heterogeneous nature of the nanoparticles, which is probably related to the fact that we are dealing with samples in which the surface to volume ratio is important. In fact, the effect of the size of the nanoparticles deserves a further study in its own right. The present study regarding the reversible switching of RT-FM offers a feasible opportunity to tailor the magnetic properties of oxides and paves the way for future investigations to develop materials.

Acknowledgements

We thank Ana M. Llois, Veronica Vildosola, Leticia Granja and Pablo Esquinazi for fruitful discussions. This work was partially supported by CONICET (PIP 00038).

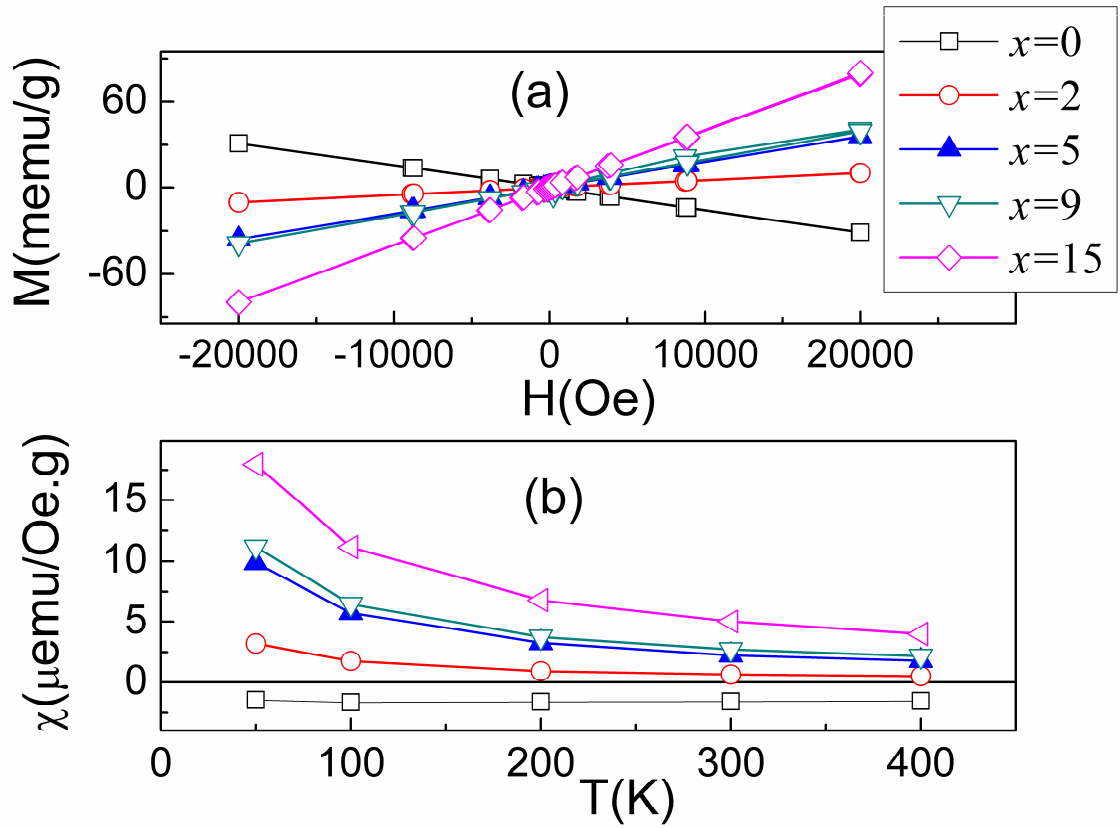

**Figure 1:** a) M vs H dependence for the as sintered samples at room temperature for different cobalt concentrations. b) Magnetic susceptibility as a function of temperature.

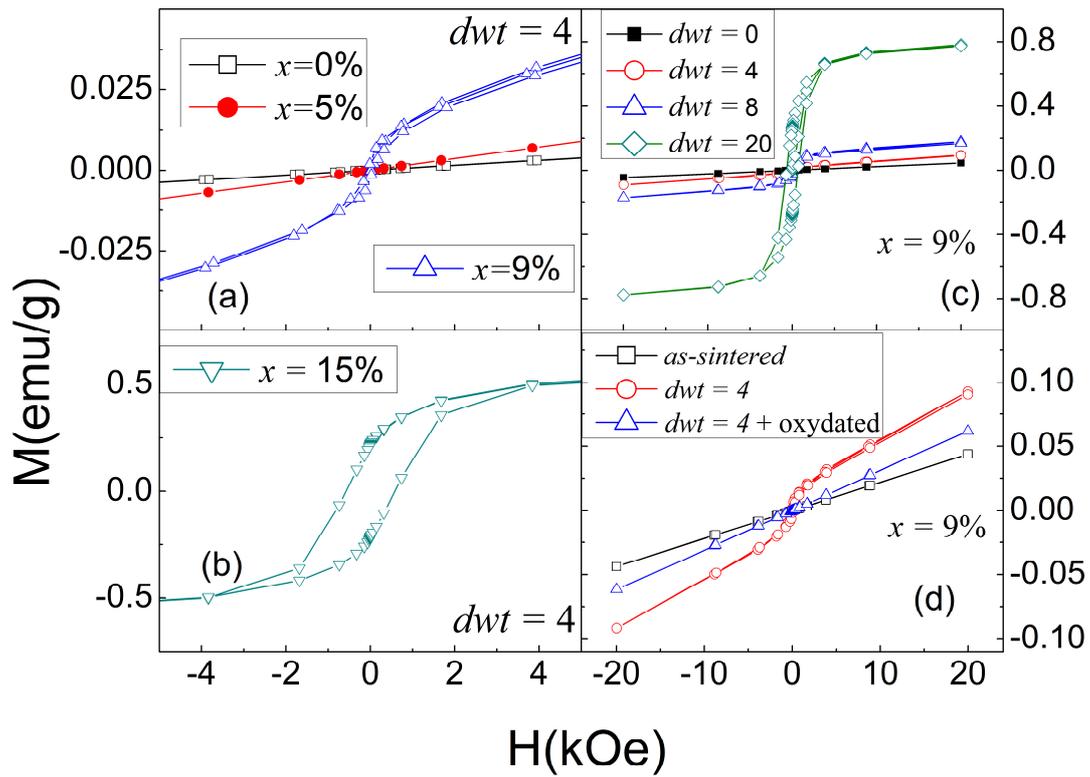

**Figure 2:** (a) and (b): M vs H dependence at 400K for samples treated in vacuum for 4 hs. (c): M vs H dependence at 400K for the samples with 9% of Co for different thermal treatments in vacuum. (d): M vs. H data for the sample with 9% of Co, measured for the as-sintered sample, the sample treated for 4 hs in vacuum and the sample re-oxydated (in air at atmospheric pressure) after been treated in vacuum for 4 hs.

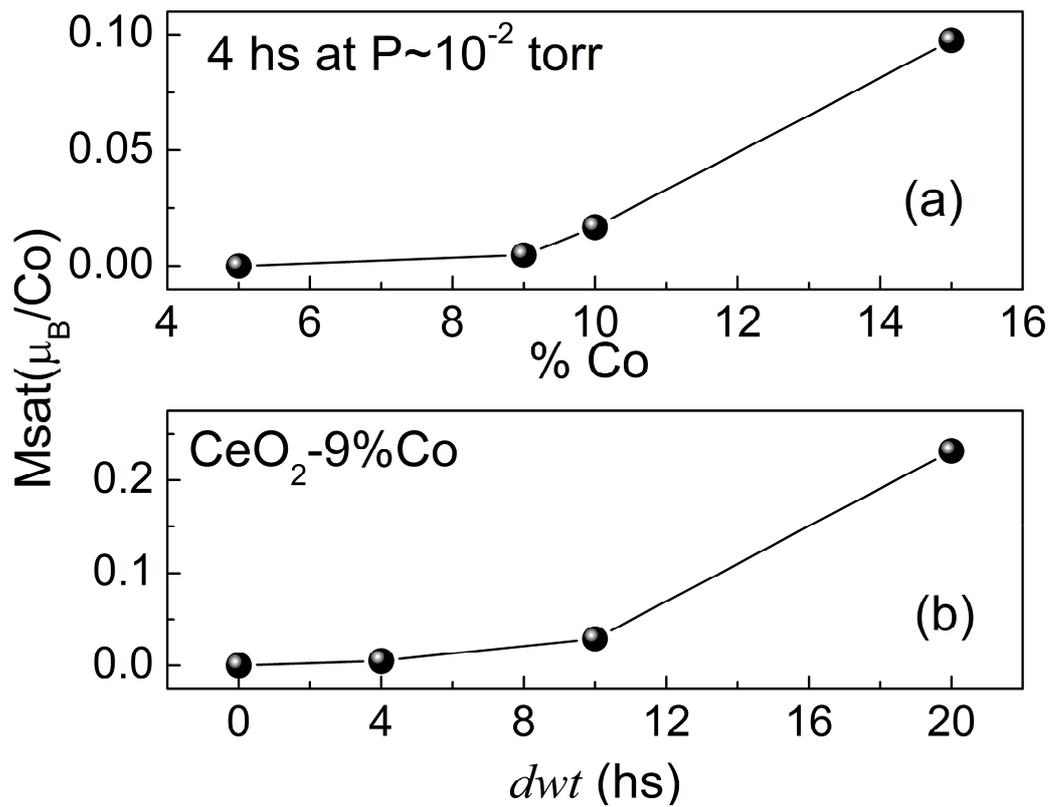

Figure 3: (a) Saturation of the FM part of the magnetization (Msat) vs. a) doping for samples treated in vacuum for 4 hs. and (b) dwell time of the treatment in vacuum (dwt).

---

## Supplementary Material

The X-ray diffraction studies were performed on the $CeO_{2-\delta}$ powder samples with nanometer grain size at the ID27 beamline of the European Synchrotron Radiation Facility using a monochromatic beam ($\lambda = 0.3738 Å$) focused to $3x2 \mu m^2$. The diffraction patterns were collected with a MAR CCD camera and the intensity vs. $2\theta$ patterns were obtained using the fit2d software[27]. The instrumental parameters (sample-detector distance, beam center and tilting angle of the detector) were refined using a $LaB_6$ powder. The samples were loaded in a 0.5mm diameter quartz capillary. A complete Rietveld refinement was done with the GSAS-EXPGUI package[27] using a cubic $F m\bar{3}m$ unit cell where scale factor, lattice parameter, peak profile and isotropic thermal factors where refined obtaining in all of the cases $\chi^2 < 0.7$ and $Rp < 5\%$. Even if we are interested in the refinement of the *O* concentration, the XRD data are not suitable for this kind of analysis therefore the *Ce* and *O* sites were considered fully occupied.

Taking the pure sample, the obtained lattice parameters are $a = (5.4138 \pm 0.0008) Å$ for the as-sintered sample, $a = (5.421 \pm 0.001) Å$ for the sample treated in vacuum for 4 hs $a = (5.4165 \pm 0.0007) Å$ for the sample treated in vacuum for 20 hs.

In figure S1 we present the diffraction data for several samples with different doping, as-sintered and reduced with different dwell time in vacuum at 500ºC. Samples were labeled with a two-number code denoting the atomic percentage of Co (*x*), and the dwell time of the vacuum treatment (*dwt*) in hours, namely *x-dwt*.

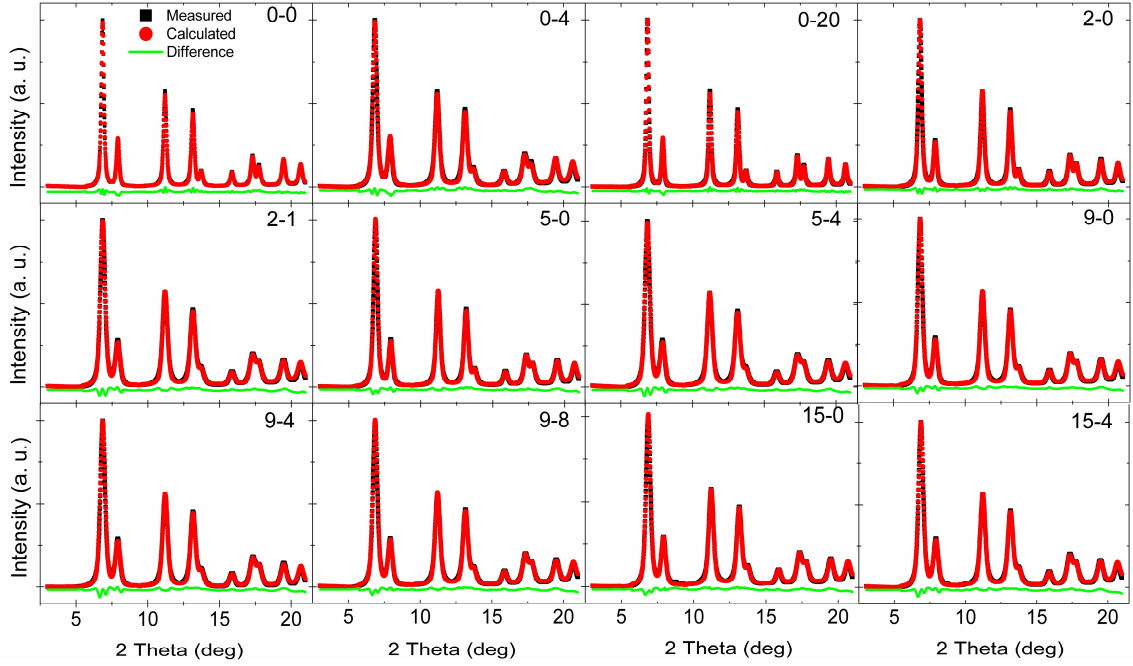

Fig. S1: X-ray diffraction data for selected samples.

The red line represents the Rietveld refinement using a cubic $Fm-3m$ structure, the black points represent the measured data and the green line is the difference between the model and the data.

We can observe the peak broadening associated with the nanometric size of the crystallites and using the Scherrer's formula ($d \approx 0.9\lambda/\beta\cos(\theta_B)$)[27] where $\lambda$ is the wavelength, $\beta$ the FWHM of the Bragg peak at $\theta_B$) we calculated a particle size $d \sim 7nm$.

We observed some impurity peaks for the nominal *Co* concentrations of 15% in the as grown samples. All the peaks can be indexed with a cubic $Co_3O_4$ phase with lattice parameter $a = (8.09 \pm 0.01)\text{Å}$ and it never overpass $1\%$ weight fraction. For the samples with thermal treatments this impurity phase are present in smaller weight fraction ($<0.2\%$) almost in the detection limit of the experiment.

In figure S2 we show the obtained parameters of the cubic lattice as a function of doping for the as-sintered samples, the samples treated in vacuum for 4 and 20 hs. In order to avoid the scattering of the data due to different grain size of each samples we calculate the lattice parameter for a 20nm sample using the refined values as it has been shown in Ref [27]. We see a monotonous reduction of the lattice parameter on doping, consistent with the substitution of $Ce^{4+}$ by smaller $Co^{2+}$ or $Co^{3+}$[27]. The non linearity observed beyond 9% of Co, suggest a solubility limit for cobalt in $CeO_2$, consistent with our finding of small impurities for the sample with 15% of Co. No clear dependence of the lattice parameter was observed with the dwell time of the thermal treatment in vacuum.

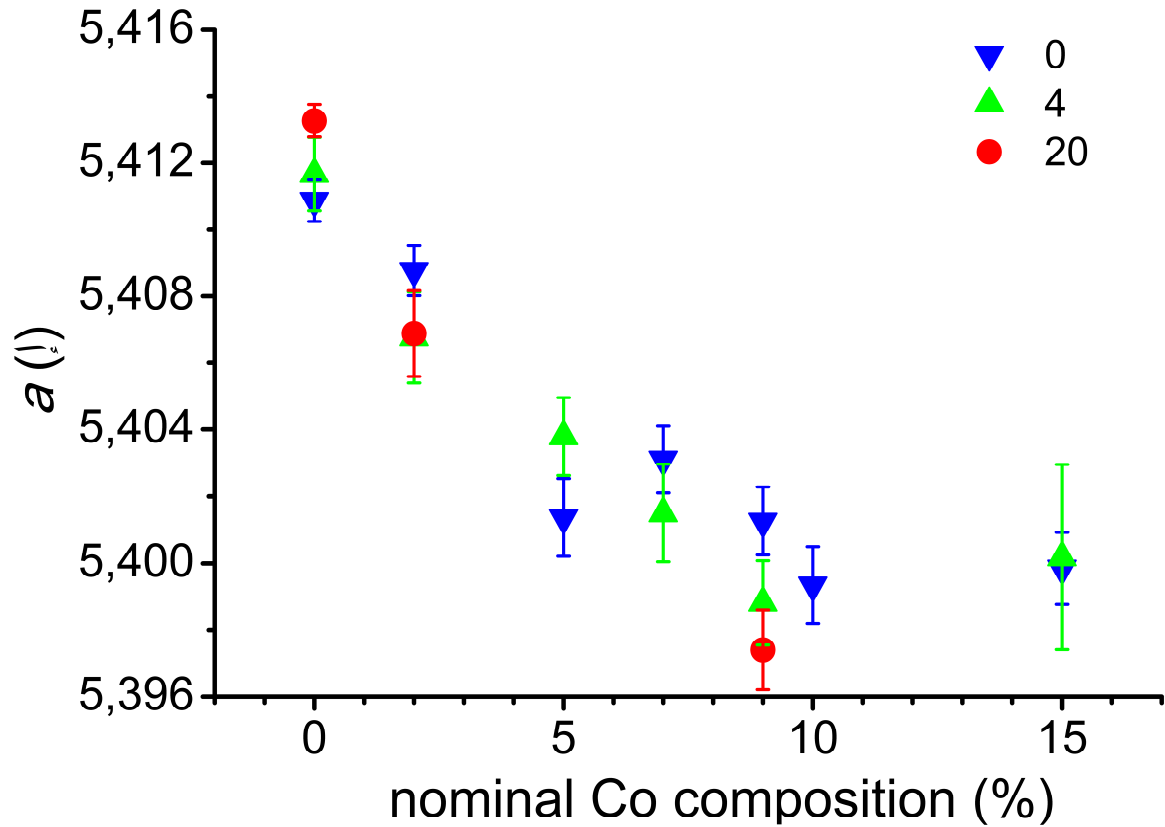

Fig S2: cubic lattice parameter as a function of doping.

In table 1 we present the number of effective Bohr Magnetons for several Co composition treated in vacuum for 4 and 8 hs. An increase is observed while increasing *dwt* except for 5% sample.

|  | Co 2% | Co 5% | Co 9% | Co 15% |
| --- | --- | --- | --- | --- |
| *dwt*=4hs | 3.4 | 4 | 3.2 | 3.3 |
| *dwt*=8hs | 4.1 | 3.8 | 4.2 | 4.1 |

**Table I: effective number of Bohr magnetons (*p*) for doped samples treated in vaccum for 4 hs and 8 hs**

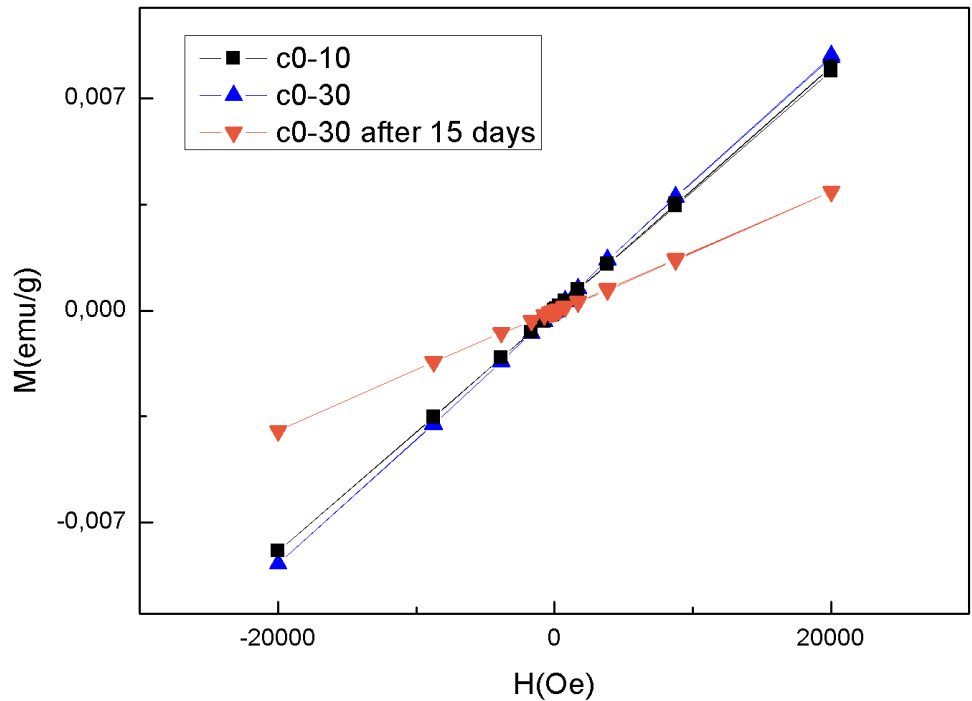

Fig. S3: Magnetization vs. magnetic field (H) at 300 K for the pure samples c0-10 and c0-30. We also show a measurement of the c0-30 sample performed 15 days after the vacuum treatment.

In figure S3 we show the dependence of the magnetization for samples reduced for 10 hs (c0-10) and 30 hs (c0-30), respectively. We see that there is no significant dependence on susceptibility between c0-30 and c0-10 samples. Also, reduced samples exhibit a time dependence. After 15 days c0-30 susceptibility decreases almost 50% as compared to that obtained just after the vacuum treatment. This behavior is probably associated with spontaneous re-oxidation in air at atmospheric pressure.

XRD resolution:

1) Detection limit: we have a background signal of around 1200counts with an absolute noise of 4 counts. The most intense Bragg peak for the sample with Co 15% has a value of 21500 counts. In this sample, we also obtained a $Co_3O_4$ peaks with a maximum intensity of 1375counts. So, we estimate less than 1% (~0.8%) of $Co_3O_4$ in this sample. Using the background noise, we can estimate that the detection limit of our X ray diffraction experiments is better than 0.1% using the "clean" 2θ–region between 8.5 and 10.4.

2) The hcp phase of Co (blue line in Fig 1) has the following reflections and intensities:

| hkl | 2θ | Rel. intensity |
|---|---|---|
| 100 | 9.88 | 23.9 |
| 002 | 10.54 | 26.5 |
| 101 | 11.2 | 100 |

while for the $Co_3O_4$ :

| hkl | 2θ | Rel. intensity |
|---|---|---|
| 311 | 8.82 | 100 |
| 222 | 9.21 | 10.6 |
| 400 | 10.64 | 20.3 |

We can clearly observe from this table that the reflections of the $Co_3O_4$ and hcp-Co in the clean 2θ–region (8.5 < 2θ < 10.4) are of the same intensity. So even if the most intense Bragg reflection of hcp-Co is overlapped with $CeO_2$ (see Fig. 1), we will be able to observe the other reflections in the clean 2θ region and we expect to have the same detection limit, i.e. ~0.05% (see Fig. 2 and Fig. 3).

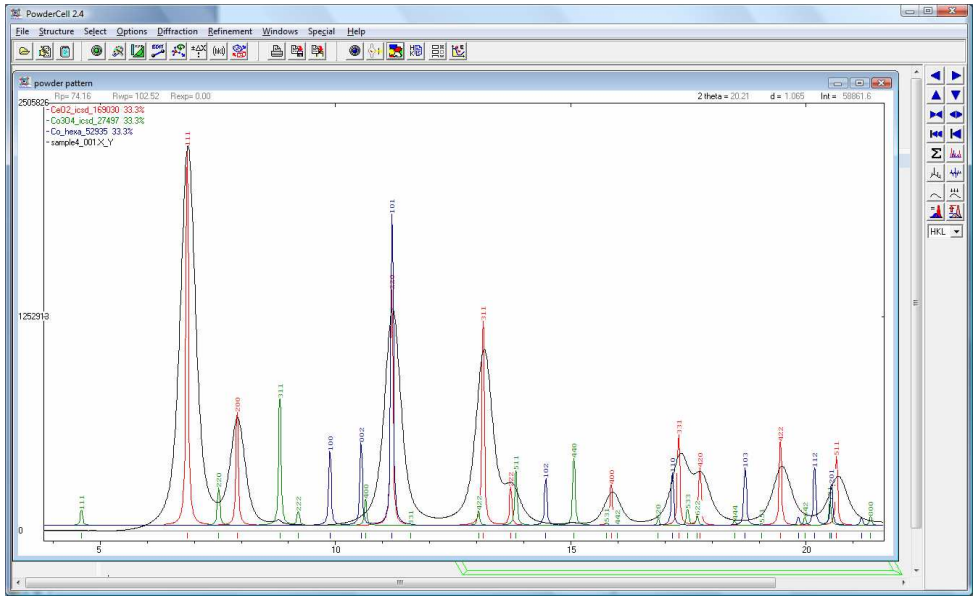

Fig. 1: Simulated diffraction pattern of $CeO_2$, $Co_3O_4$ and Co-hcp in color lines together with the experimental data of the sample $CeO_2$ with 15% of Cobalt (black line). The phase intensities are simulated just to show the position of the Bragg reflections.

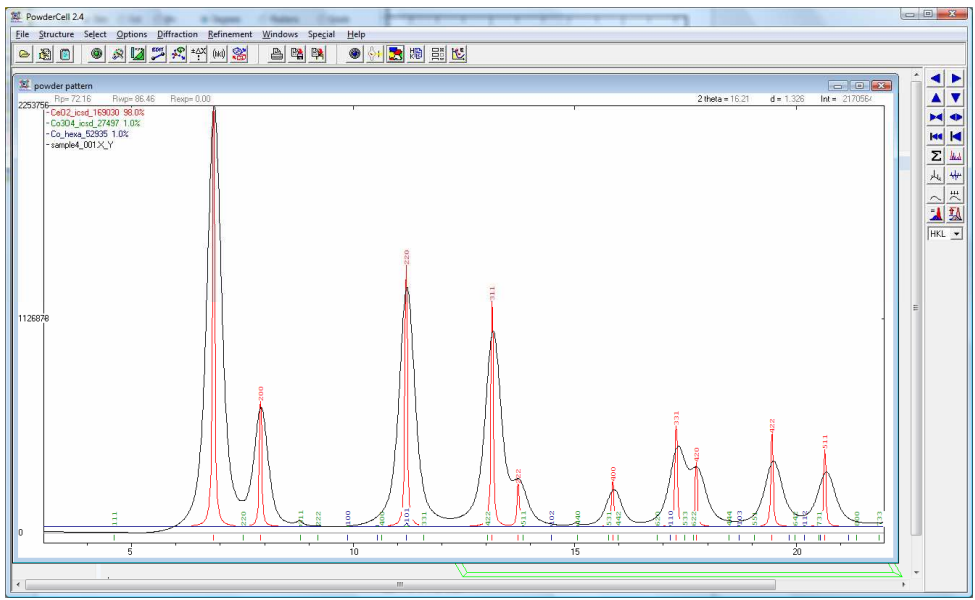

Fig. 2: Simulated diffraction pattern of $CeO_2$, $Co_3O_4$ and Co-hcp in color lines together with the experimental data (black line). The phase intensities are representative of the estimated phase concentrations for the sample $CeO_2$ with 15% of Co.

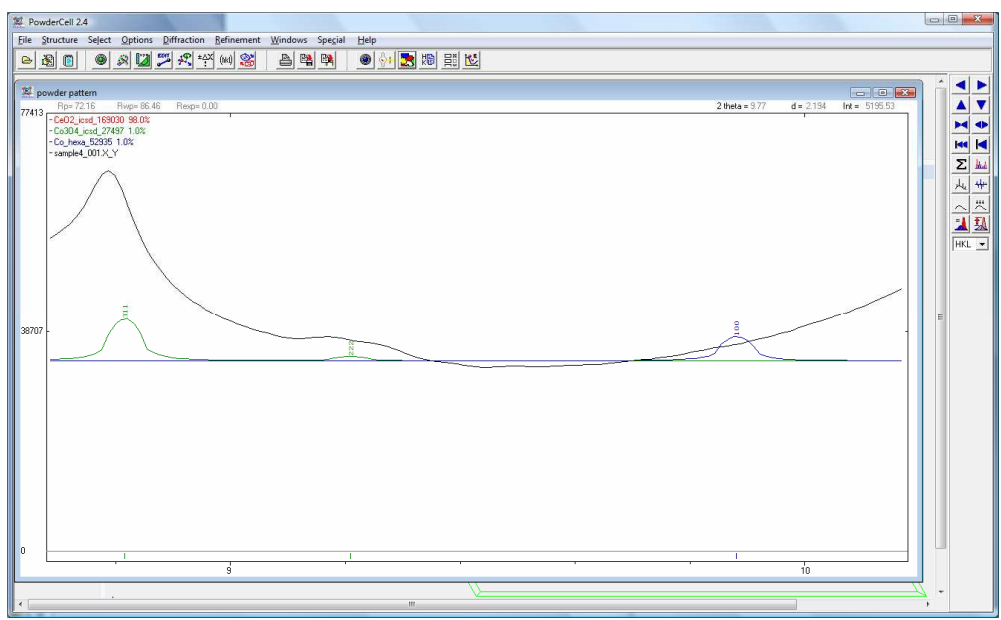

Fig. 3: Zoom of the "clean 2θ region" (8.5<2θ<10.4) of the simulated diffraction pattern of $CeO_2$, $Co_3O_4$ and Co-hcp in color lines together with the experimental data (black line). The phase intensities are representative of the estimated phase concentrations for the sample $CeO_2$ with 15% of Cobalt.